\documentclass[]{interact}

\usepackage{epstopdf}
\usepackage[caption=false]{subfig}

\usepackage[longnamesfirst,sort]{natbib}
\usepackage{amssymb}

\usepackage{graphicx}%
\usepackage{amsmath}
\usepackage{amsthm}
\newtheorem{assumption}{Assumption}
\usepackage{amssymb,amsmath,amsthm}

\usepackage{multirow}%
\usepackage{amsmath,amssymb,amsfonts}%
\usepackage{amsthm}%
\usepackage{mathrsfs}%
\usepackage[title]{appendix}%
\usepackage{hyperref}
\usepackage{manyfoot}%
\usepackage{booktabs}%
\usepackage{algorithm}
\usepackage{algorithmic}
\usepackage{listings}%
\usepackage[utf8]{inputenc} 
\usepackage[T1]{fontenc}    
\usepackage{hyperref}       
\usepackage{url}            
\usepackage{booktabs}       
\usepackage{amsfonts}       
\usepackage{nicefrac}       
\usepackage{rotating}
\usepackage{savesym}
\usepackage{graphicx}
\usepackage{adjustbox}
\usepackage{xcolor}
\savesymbol{AND}
\usepackage{epstopdf}
\usepackage{adjustbox}
\usepackage{tabularx}
\usepackage{lscape}

\usepackage{mathtools}
\usepackage{longtable,pdflscape,booktabs}
\DeclarePairedDelimiterX{\norm}[1]{\lVert}{\rVert}{#1}
\DeclarePairedDelimiterX{\inp}[2]{\langle}{\rangle}{#1, #2}
\usepackage{rotating}
\usepackage{longtable}
\usepackage{adjustbox}
\usepackage{booktabs} 
\usepackage{comment}

\begin{document}

\articletype{Original Research Article}

\title{Copula-Based Endogeneity Correction for Doubly Robust Estimation of Treatment Effect}

\author{
\name{ Sahil Shikalgar\textsuperscript{a}  and Md. Noor-E-Alam\textsuperscript{a}\thanks{CONTACT Md. Noor-E-Alam. Email: mnalam@neu.edu}}
\affil{\textsuperscript{a}Department of Mechanical and Industrial Engineering, Northeastern University, Boston, MA 02115, USA}
}

\maketitle

\begin{abstract}

 Doubly Robust (DR) estimation of treatment effect relies on an untestable assumption that is the absence of unobserved confounding. This assumption is particularly problematic in the context of healthcare research, where variables like prescription refill rates serve as proxies for unobserved behaviors such as medication adherence. These proxy variables are often endogenous, exhibiting correlation with the regression error term due to unmeasured confounding or measurement error. We propose a copula-corrected doubly robust estimator that addresses endogeneity in both the treatment and outcome models without requiring instrumental variables. Gaussian copulas model the joint distribution of endogenous covariates and the error term, enabling consistent estimation while preserving the doubly robust property that requires correct specification of either the treatment or outcome model, not both. Monte Carlo simulations demonstrate that naive DR estimation exhibits substantial bias under endogeneity, whereas our corrected estimator recovers unbiased treatment effects across different data-generating processes. We apply our method to examine the effect of nutritional counseling on blood pressure using National Health and Nutrition Examination Survey (NHANES) data. Naive DR estimation suggests counseling is associated with increased blood pressure. After copula correction, this effect becomes statistically insignificant, consistent with literature showing modest effects of nutritional counseling in reducing blood pressure. Our methodology provides researchers with a practical tool for obtaining treatment effects in presence of endogeneity.

\end{abstract}

\begin{keywords}
Doubly Robust Estimation; Endogeneity; Copula Correction; Causal Inference; Observational Studies; Nutrition Counseling 
\end{keywords}

\section{Introduction}\label{sec:Intro}

Causal inference from observational data requires accurate measurement of confounding variables that influence treatment selection and outcomes. This requirement is rarely satisfied in practice, as variables like medication adherence, and health literacy are often unobservable. Researchers instead rely on proxy variables \citep{wyss2022machine} like prescription refill rates to approximate adherence behavior, and education level  for health literacy. However, these proxies  fail to capture complete information \citep{seltzer2021perilous}. The unmeasured component of proxy variables violates the unconfoundedness assumption \citep{rosenbaum1983central, imbens2015causal}, as treatment assignment is no longer independent of potential outcomes conditional on observed covariates alone. This incomplete measurement presents itself as endogeneity \citep{villas1999endogeneity}, a correlation between observed covariates and the regression error term. Consider a study estimating the effect of antihypertensive medication on blood pressure using prescription refill rates as a proxy for medication adherence. Patients who consistently refill prescriptions likely to differ  from inconsistent refill patients. They may engage in dietary modifications, exercise regularly, monitor their blood pressure at home, or possess greater health consciousness. Standard methods like, doubly robust estimation \citep{funk2011doubly}, attribute endogeneity effects to the treatment itself leading to biased results. Addressing this challenge has motivated several methodological approaches, each with distinct advantages and limitations.

The canonical solution to endogeneity is to use instrumental variables (IV) \citep{angrist2001instrumental}. 
However, this approach faces two critical challenges in healthcare applications. First, the exclusion restriction is fundamentally untestable \citep{van2007economic, mellon2025rain} and secondly, credible instruments satisfying this restriction are difficult to find  \citep{brookhart2006evaluating}. An alternative approach uses sensitivity analysis \citep{vanderweele2011bias,cinelli2020making} to assess how robust causal estimates are to potential violations of the unconfoundedness assumption. While valuable for assessing robustness, these methods do not correct for endogeneity, they simply bound its impact. 

More recently, copula-based methods have emerged as instrument-free alternatives to address endogeneity through distributional assumptions rather than exclusion restrictions \citep{park2012handling}. Copulas model the joint distribution of endogenous covariates and the error term by separating the marginal distributions from their dependence structure. This allows researchers to recover consistent estimates by explicitly modeling the correlation induced by unmeasured confounding \citep{smith2003modelling}. Copula methods exploit the information contained in the endogenous variable's distribution to correct for endogeneity. Unlike the exclusion restriction, these assumptions particularly the non-normality of endogenous variables can be assessed through diagnostic tests \citep{park2012handling}.  This approach has been valuable where valid IVs not available.

Despite these advantages, existing copula-based methods share a critical limitation with parametric methods, they require correct specification of the treatment assignment model or the outcome regression model to achieve consistent estimates. 
Doubly robust (DR) estimation offers a solution to model uncertainty in causal inference by combining the propensity score model and the outcome regression model in a way that yields consistent estimates if either model is correct \citep{robins1994estimation}. This protection against misspecification has made DR methods popular in observational healthcare research. However, DR estimator rely on the unconfoundedness assumption \citep{funk2011doubly}. When covariates are endogenous, as with the proxy variables discussed earlier naive DR estimation is biased.
To our knowledge, no existing work addresses both challenges simultaneously. Copula methods handle endogeneity but lack robustness to model misspecification. DR methods provide protection against misspecification but assume exogenous covariates. This gap is particularly relevant given that both problems  (endogeneity and model uncertainty) are important considerations in observational healthcare studies.

This paper develops a Copula based Endogeneity corrected Doubly Robust estimator (CEDR). CEDR integrates gaussian copula corrections for endogenous covariates into the doubly robust framework, yielding consistent treatment effect estimates. The method requires no instrumental variables, instead exploits non-normality in endogenous covariates for identification. Providing a comprehensive solution to the  endogeneity and model misspecification problem.

The remainder of this paper proceeds as follows. Section 2 reviews the doubly robust estimation framework and introduces our copula-corrected estimator. Section 3 presents Monte Carlo simulation demonstrating that naive doubly robust estimation exhibits substantial bias when covariates are endogenous, while our corrected estimator recovers unbiased treatment effects. Section 4 applies our method to estimate the effect of nutrition counseling on blood pressure using National Health and Nutrition Examination Survey (NHANES) data \citep{zipf2013health}. Section 5 includes a discussion of our findings, and Section
6 concludes the paper with key takeaways and future directions.

\section{Methodology}\label{sec:Method}

This section establishes the formal framework for our proposed estimator. We begin with notation and the causal inference setup, then review DR estimation and Gaussian copula methods for endogeneity correction. Building on these foundations, we then introduce the CEDR estimator.

\subsection{Setup and Notation}\label{sec:2.1}
In this work we adopt the potential outcomes framework \citep{rubin1974estimating, holland1986statistics}. Consider a sample of $n$ units  ($i = 1, \dots, n$). Each unit receives a binary treatment $T_i \in \{0,1\}$ and has an observed outcome $Y_i$. Under the Stable Unit Treatment Value Assumption (SUTVA), which requires that one unit's treatment assignment does not affect another unit's outcome and that treatments are consistent. $Y_i(1)$ is the outcome under treatment and $Y_i(0)$ under control as shown in equation \ref{eq:consistency}. 

\begin{equation}\label{eq:consistency}
    Y_i = T_i \, Y_i(1) + (1 - T_i) \, Y_i(0)
\end{equation}

 Let $\mathbf{X}_i \in \mathbb{R}^p$ denote a vector of observed pre-treatment covariates that may be related to treatment and/or the outcome. 

\begin{assumption}[Unconfoundedness]\label{ass:unconf}
    $\{Y_i(1),\, Y_i(0)\} \perp\!\!\!\perp T_i \mid \mathbf{X}_i$.
\end{assumption}

\noindent Conditional on observed covariates, treatment assignment is independent of potential outcomes.

\begin{assumption}[Overlap]\label{ass:overlap}
    $0 < P(T_i=1 | \mathbf{X}_i) < 1 $.
\end{assumption}

\noindent Every unit has a positive probability of receiving treatment.

\subsection{Naive Doubly Robust Estimator}\label{sec:2.2}

The DR estimator, originally proposed by \citet{robins1994estimation}, combines outcome regression with inverse probability weighting (IPW) to estimate causal effects. This approach yields consistent estimates of the Average Treatment Effect (ATE) if either the propensity score model or the outcome regression model is correctly specified, but not necessarily both \citep{funk2011doubly, bang2005doubly}.

We define the propensity score model as $e(\mathbf{X}_i) = P(T_i = 1 \mid \mathbf{X}_i)$ and the outcome regression models $m_1(\mathbf{X}_i, \alpha_1)$ and $m_0(\mathbf{X}_i, \alpha_0)$, where $m_t(\mathbf{X}_i, \alpha_t)$ is a parametric model for $E[Y_i \mid T_i = t, \mathbf{X}_i]$ where $t \in \{0, 1\}$. The DR estimator of the ATE (${\tau}_{DR}$) takes the form:
\begin{align}\label{eq:dr}
\begin{split}
    \hat{\tau}_{DR} = \frac{1}{n} \sum_{i=1}^{n} \left[ \frac{T_i Y_i}{e(\mathbf{X}_i, \hat{\beta})} - \frac{T_i - e(\mathbf{X}_i, \hat{\beta})}{e(\mathbf{X}_i, \hat{\beta})} \, m_1(\mathbf{X}_i, \hat{\alpha}_1) \right] \\
    - \frac{1}{n} \sum_{i=1}^{n} \left[ \frac{(1 - T_i) Y_i}{1 - e(\mathbf{X}_i, \hat{\beta})} + \frac{T_i - e(\mathbf{X}_i, \hat{\beta})}{1 - e(\mathbf{X}_i, \hat{\beta})} \, m_0(\mathbf{X}_i, \hat{\alpha}_0) \right]
\end{split}
\end{align}
where $\hat{\beta}$, $\hat{\alpha}_1$, and $\hat{\alpha}_0$ are estimated from the propensity score and outcome regression models. The first terms in each average are inverse probability weighted estimators for $E[Y_i(1)]$ and $E[Y_i(0)]$. The second terms serve as augmentation corrections that provide the robustness property.

However, the double robustness property relies on Assumption~\ref{ass:unconf}. Suppose $\mathbf{X}^{endo} \subset \mathbf{X}$ are endogenous, that is, correlated with the structural error terms. Then both $e(\mathbf{X}_i)$ and $m_t(\mathbf{X}_i, \alpha_t)$ are estimated from incomplete information.
 In this case, correctly specifying either model is insufficient because the bias originates not from misspecification but from the endogeneity of the conditioning variables themselves,  violating Assumption 1.

\subsection{Gaussian Copula Correction for Endogenous covariates}\label{sec:2.3}

When valid IVs are unavailable, copula-based methods offer an alternative approach to addressing endogeneity by exploiting distributional assumptions \citep{park2012handling}. The key insight is that if the joint distribution of an endogenous covariate and the regression error term can be modeled, the endogeneity problem becomes tractable.

Consider a model with an endogenous covariate $X^{endo}$ whose correlation with the error term $\varepsilon$ arises from unmeasured confounding:
\begin{equation}\label{eq:outcome_endo}
    Y_i = \alpha + \beta X_i^{endo} + \varepsilon_i
\end{equation}
where $\beta$ is the coefficient of interest and $X_i^{endo}$ has a strictly monotonic CDF and is correlated with $\varepsilon_i$. The Gaussian copula approach addresses this endogeneity using three assumptions \citep{park2012handling, liengaard2025dealing}:
\begin{assumption}\label{assu:nonnormal}
The endogenous covariate $X_i^{\text{endo}}$ is non-normally distributed.
\end{assumption}
\begin{assumption}\label{assu:copula}
The dependence between $X_i^{\text{endo}}$ and $\varepsilon_i$ follows a Gaussian copula structure:
\begin{equation}\label{eq:copula_structure}
    \begin{pmatrix} \varepsilon_i \\ X_i^* \end{pmatrix} \sim N\left( \begin{bmatrix} 0 \\ 0 \end{bmatrix}, \begin{bmatrix} \sigma^2_\varepsilon & \rho\sigma_\varepsilon \\ \rho\sigma_\varepsilon & 1 \end{bmatrix} \right)
\end{equation}
where $X_i^* = \Phi^{-1}(F_X(X_i^{\text{endo}}))$, $F_X(\cdot)$ is the marginal CDF of $X_i^{\text{endo}}$, $\Phi^{-1}(\cdot)$ is the inverse standard normal CDF, and $\rho$ captures the degree of endogeneity.
\end{assumption}
\begin{assumption}\label{assu:normalerror}
The structural error is normally distributed, $\varepsilon_i \sim N(0, \sigma^2_\varepsilon)$ with $\sigma^2_\varepsilon > 0$.
\end{assumption}

Under these assumptions, \citet{park2012handling} show that the Gaussian copula approach provides an unbiased, IV-free estimator of $\beta$ using control functions. The copula term $C(X_i^{\text{endo}}) = \Phi^{-1}(F_X(X_i^{\text{endo}}))$ acts as a control function that absorbs the correlation between $X_i^{\text{endo}}$ and $\varepsilon_i$. Adding this term to Equation~\eqref{eq:outcome_endo} yields the augmented regression:
\begin{equation}\label{eq:copula_augmented}
    \widetilde{Y_i} = \alpha + \beta X_i^{\text{endo}} + \gamma \, C(X_i^{\text{endo}}) + u_i
\end{equation}
where $\gamma =\rho \sigma_\varepsilon $ and $u_i$ is uncorrelated with all regressors, allowing consistent estimation of $\beta$ via ordinary least squares or maximum likelihood.

In practice, $F_X(\cdot)$ is unknown and must be estimated from the data. Following \citet{liengaard2025dealing}, we employ an adjusted empirical cumulative distribution function (ECDF):
\begin{equation}\label{eq:adjusted_ecdf}
    \hat{F}_X(x) = \frac{1}{2n} + \frac{n-1}{n^2} \sum_{i=1}^{n} \mathbf{I}(X_i \leq x)
\end{equation}
which avoids the boundary problem where $\hat{F}_X(X_i) = 0$ or $1$ would produce infinite values under $\Phi^{-1}(\cdot)$. The estimated copula term is then $\hat{C}(X_i^{endo}) = \Phi^{-1}(\hat{F}_X(X_i^{endo}))$.

Identification requires that $X_i^{endo}$ be non-normally distributed. When $X_i^{endo}$ is normal, $\Phi^{-1}(F_X(X_i))$ reduces to a linear transformation of $X_i^{endo}$, creating perfect collinearity and rendering $\beta$ and $\gamma$ unidentifiable \citep{park2012handling}. However, unlike the exclusion restriction, the non-normality can be empirically assessed through diagnostic tests.

\subsection{CEDR Estimator}\label{sec:2.4}

Recall from Section~\ref{sec:2.2} that the standard DR estimator requires Assumption~\ref{ass:unconf} that $\mathbf{X}_i$ captures all common causes of treatment and outcome. When covariates are endogenous, this assumption is violated. As established in Section~\ref{sec:2.3}, the Gaussian copula correction addresses covariate endogeneity by eliminating the correlation between observed covariates and the structural error. The CEDR estimator introduces this correction in the DR framework.

\subsubsection{Setup}

Partition the covariate vector $\mathbf{X}_i$ into exogenous components $\mathbf{X}_i^{{exo}}$ and endogenous components $\mathbf{X}_i^{{endo}}$. For each endogenous covariate $X_{i}^{{endo}_j}$, $j = 1, \dots, q$, construct the copula correction term:
\begin{equation}\label{eq:copula_term}
    C_j(X_{i}^{{endo}_j}) = \Phi^{-1}\!\left(\hat{F}_{X}(X_{i}^{{endo}_j})\right)
\end{equation}
where $\hat{F}_{X}(\cdot)$ is the adjusted ECDF from Equation~\eqref{eq:adjusted_ecdf}. Let $\mathbf{C}_i = \bigl(C_1(X_{i}^{{endo}_1}), \dots, C_q(X_{i}^{{endo}_q})\bigr)$ denote the vector of copula correction terms. Define the updated covariate vector as:
\begin{equation}\label{eq:augmented_covariates}
    \widetilde{\mathbf{X}}_i = \bigl(\mathbf{X}_i,\; \mathbf{C}_i\bigr)
\end{equation}

\subsubsection{Corrected Models} 

The CEDR estimator proceeds in two steps.

\textit{Step 1: Corrected model estimation.} Estimate the propensity score and outcome regression models using the augmented covariates $\widetilde{\mathbf{X}}_i$. Since the Gaussian copula framework assumes normally distributed errors (Assumption~\ref{assu:normalerror}), we use a probit (instead of logit) for the propensity score model:
\begin{align}
    {e}(\widetilde{\mathbf{X}}_i, \hat{\beta}) &= P(T_i = 1 \mid \widetilde{\mathbf{X}}_i) \label{eq:corrected_ps} \\[4pt]
    {m}_t(\widetilde{\mathbf{X}}_i, \hat{\alpha}_t) &= E[Y_i \mid T_i = t, \widetilde{\mathbf{X}}_i], \quad t \in \{0,1\} \label{eq:corrected_outcome}
\end{align}
 $\mathbf{C}_i$ in the conditioning set absorbs the correlation between endogenous covariates and the structural error. Importantly, the copula terms are included in the model estimation to correct coefficient bias, but are excluded from the prediction step.

\textit{Step 2: Doubly robust estimation.} The corrected models in Equations~\eqref{eq:corrected_ps}--\eqref{eq:corrected_outcome} yield coefficient estimates on three components: $\mathbf{X}_i^{\text{exo}}$, $\mathbf{X}_i^{\text{endo}}$, and the copula terms $\mathbf{C}_i$. For ATE estimation, we retain only the debiased coefficients on $\mathbf{X}_i^{\text{exo}}$ and $\mathbf{X}_i^{\text{endo}}$ and discard $\mathbf{C}_i$. Let $\hat{\beta}_{\mathbf{X}}$ and $\hat{\alpha}_{t,\mathbf{X}}$ denote the retained coefficients:
\begin{align}
    \hat{e}_{\text{CEDR}} &= {e}({\mathbf{X}}_i,\hat{\beta}_{\mathbf{X}}) \label{eq:cedr_ps} \\[4pt]
    \hat{m}_{t,\text{CEDR}} &= {m}_t({\mathbf{X}}_i, \hat{\alpha}_{t,\mathbf{X}}), \quad t \in \{0,1\} \label{eq:cedr_outcome}
\end{align}
where $\mathbf{X}_i = (\mathbf{X}_i^{\text{exo}}, \mathbf{X}_i^{\text{endo}})$.  The CEDR estimate of the ATE is then obtained by substituting Equations~\eqref{eq:cedr_ps}--\eqref{eq:cedr_outcome} into the doubly robust estimator in Equation~\eqref{eq:dr}:

\begin{align}\label{eq:cedr_ate}
\begin{split}
    \hat{\tau}_{\text{CEDR}} = \frac{1}{n} \sum_{i=1}^{n} \left[ \frac{T_i Y_i}{\hat{e}_{\text{CEDR}}} - \frac{T_i - \hat{e}_{\text{CEDR}}}{\hat{e}_{\text{CEDR}}} \, \hat{m}_{1,\text{CEDR}}\right] \\
    - \frac{1}{n} \sum_{i=1}^{n} \left[ \frac{(1 - T_i) Y_i}{1 - \hat{e}_{\text{CEDR}}} + \frac{T_i - \hat{e}_{\text{CEDR}}}{1 - \hat{e}_{\text{CEDR}}} \, \hat{m}_{0,\text{CEDR}}\right]
\end{split}
\end{align}

\subsection{Practical Implementation}\label{sec:2.5}
This section provides guidance on when and how to apply the CEDR estimator in practice.

\subsubsection{When to Use the CEDR Estimator}

The CEDR estimator is designed for observational studies in which three conditions hold simultaneously (i) proxy covariates are used, inducing covariate endogeneity (ii) credible IVs satisfying the exclusion restriction are unavailable and (iii) there is uncertainty about the correct functional form of either the propensity score or the outcome model.

\subsubsection{Non-normality Checks}
 Identification requires that the endogenous covariates are non-normal. This can be assessed through visual inspection (histograms, Q-Q plots) and formal tests such as the Anderson-Darling or Cram\'{e}r-von Mises tests \citep{park2012handling, falkenstrom2023using}.

\subsubsection{Inference}

Since the copula correction terms are generated regressors, the analytic standard errors from the corrected regression may underestimate the true variability. We recommend using bootstrap to obtain valid standard errors and confidence intervals.

\begin{algorithm}[h]
\caption{CEDR Estimation Procedure}\label{alg:cedr}
\begin{algorithmic}[1]
\REQUIRE Observed data $\{(Y_i, T_i, \mathbf{X}_i)\}_{i=1}^n$; partition $\mathbf{X}_i$ into $\mathbf{X}_i^{\text{exo}}$ and $\mathbf{X}_i^{\text{endo}}$
\ENSURE Estimated ATE $\hat{\tau}_{\text{CEDR}}$
\STATE \textbf{Verify assumptions:} Test non-normality of $\mathbf{X}_i^{\text{endo}}$
\FOR{each endogenous covariate $X_i^{{\text{endo}}_j}$, $j = 1, \dots, q$}
    \STATE Estimate $\hat{F}_{X^{}}(\cdot)$ using the adjusted ECDF (Equation~\eqref{eq:adjusted_ecdf}).
    \STATE Compute copula term $C_j(X_i^{{\text{endo}}_j}) = \Phi^{-1}\!\left(\hat{F}_{X^{}}(X_i^{{\text{endo}}_j})\right)$.
\ENDFOR
\STATE Form augmented covariates $\widetilde{\mathbf{X}}_i = (\mathbf{X}_i, \mathbf{C}_i)$.
\STATE Estimate corrected propensity score $\hat{e}(\widetilde{\mathbf{X}}_i, \hat{\beta})$ via Equation~\eqref{eq:corrected_ps}.
\STATE Estimate corrected outcome models $\hat{m}_t(\widetilde{\mathbf{X}}_i, \hat{\alpha}_t)$ for $t \in \{0,1\}$ via Equation~\eqref{eq:corrected_outcome}.
\STATE Compute $\hat{\tau}_{\text{CEDR}}$ via Equation~\eqref{eq:cedr_ate}.
\STATE Obtain bootstrap standard errors and confidence intervals by resampling Steps 2--9.
\end{algorithmic}
\end{algorithm}

\section{Monte Carlo Simulations}\label{sec:simulations}

In this section, we evaluate the performance of the CEDR estimator against the naive doubly robust (DR) estimator through Monte Carlo simulations. We assess bias in the estimated average treatment effect, the doubly robust property under model misspecification.

\subsection{Simulation Design}\label{sec:sim_design}

In the data-generating process an endogenous covariate is correlated with the error terms in both the outcome and treatment equations, mimicking a proxy variable. The parameter $\rho$ controls the strength of endogeneity. We consider two designs, scenario 1 with one endogenous covariate and three observed covariates, and an scenario 2 with two endogenous covariates and six observed covariates.

\subsubsection{Scenario 1}\label{sec:sim_baseline}

We generate $n$ observations with observed covariates ($Z_1, Z_2, Z_3$), of which $Z_1$ is endogenous.

\begin{equation}\label{eq:sim_latent}
    \begin{pmatrix} \varepsilon_i \\ Z_{1i}^* \\ \upsilon_i \end{pmatrix} \sim N\!\left(\begin{bmatrix} 0 \\ 0 \\ 0 \end{bmatrix},\; \begin{bmatrix} 1 & \rho & 0 \\ \rho & 1 & \rho \\ 0 & \rho & 1 \end{bmatrix}\right)
\end{equation}
where $\varepsilon_i$ is the error term in the outcome equation, $Z_{1i}^*$ is the latent covariate, and $\upsilon_i$ is the error term in the treatment equation. The parameter $\rho$ controls how strongly $Z_{1i}^*$ correlates with both errors.

The endogenous covariate $Z_{1i} \sim \chi^2(3)$ (standardized) is obtained via the probability integral transform of $Z_{1i}^*$, inducing the non-normality. The exogenous covariates are $Z_{2i} \sim N(0,1)$ and $Z_{3i} \sim \text{Bernoulli}(0.3)$, both independent of $\varepsilon_i$ and $\upsilon_i$.

The treatment model is generated as:
\begin{equation}\label{eq:sim_ps}
    P(T_i = 1 \mid Z_{1i}, Z_{2i}, Z_{3i}) = \Phi(\gamma_0 + Z_{1i} - Z_{2i} + Z_{3i})
\end{equation}
where $\Phi(\cdot)$ is the standard normal CDF, and the intercept $\gamma_0$ is calibrated to achieve $P(T_i = 1) \approx 0.30$. 

The outcome model is generated as:
\begin{equation}\label{eq:sim_outcome}
    Y_i = Z_{1i} + Z_{3i} + \tau \cdot T_i + \varepsilon_i
\end{equation}
where $\tau = 2$ is the true average treatment effect.

\subsubsection{Scenario 2}\label{sec:sim_extended}
The extended design adds a endogenous covariate $Z_{4i}$ and two exogenous covariates $Z_{5i}$ and $Z_{6i}$. 
\begin{equation}\label{eq:sim_latent_ext}
\begin{pmatrix} \varepsilon_i \\ Z_{1i}^* \\ Z_{4i}^* \\ \upsilon_i \end{pmatrix} \sim N\!\left(\begin{bmatrix} 0 \\ 0 \\ 0 \\ 0 \end{bmatrix}, \begin{bmatrix} 1 & \rho & \rho & 0 \\ \rho & 1 & 0 & \rho \\ \rho & 0 & 1 & \rho \\ 0 & \rho & \rho & 1 \end{bmatrix}\right)
\end{equation}
Both $Z_{1i}$ and $Z_{4i} \sim \chi^2(3)$ (standardized) are obtained via the probability integral transform of $Z_{1i}^*$ and $Z_{4i}^*$, respectively. The exogenous covariates are $Z_{2i}, Z_{5i}, Z_{6i} \sim N(0,1)$ and $Z_{3i} \sim \text{Bernoulli}(0.3)$, all independent of $\varepsilon_i$  and $\upsilon_i$. 

The treatment model is generated as:
\begin{equation}\label{eq:sim_ps_ext}
\begin{aligned}
    P(T_i = 1 \mid \mathbf{Z}_i) = \Phi\big( & \gamma_0 + Z_{1i} - 2Z_{2i} + Z_{3i} \\
    & + Z_{4i} - 2Z_{5i} + Z_{6i}\big)
\end{aligned}
\end{equation}
where $\gamma_0$ is calibrated to achieve $P(T_i = 1) \approx 0.30$.

The outcome is generated as:
\begin{equation}\label{eq:sim_outcome_ext}
    Y_i = Z_{1i} + Z_{3i} + Z_{4i} + Z_{6i} + \tau \cdot T_i + \varepsilon_i
\end{equation}
where $\tau = 2$ is the true treatment effect. 

\subsubsection{Misspecification Scenarios}\label{sec:sim_scenarios}

For each design, we evaluate three specifications to test the doubly robust property. Misspecification is induced by omitting $Z_{3i}$ from the model:

\begin{enumerate}
    \item \textbf{Both Correct:} Both the propensity score and outcome models include all observed covariates.
    \item \textbf{PS Wrong:} The propensity score omits $Z_{3i}$; the outcome model is correct.
    \item \textbf{Outcome Wrong:} The outcome model omits $Z_{3i}$; the propensity score is correct.
\end{enumerate}

Under the doubly robust property, Scenarios 2 and 3 should still yield consistent estimates when one model is correct. The CEDR estimator should preserve this property even under endogeneity, whereas the naive DR estimator is expected to be biased whenever $\rho > 0$.

\subsubsection{Experimental Conditions}\label{sec:sim_conditions}

We vary the following parameters:
\begin{itemize}
    \item \textbf{Sample size:} $n \in \{ 2000, 4000, 8000, 16000, 32000\}$
    \item \textbf{Endogeneity strength:} $\rho \in \{0.0, 0.3, 0.5\}$
\end{itemize}

\subsection{Simulation Results}\label{sec:sim_results}

\begin{figure*}[t]
    \centering
    \includegraphics[width=\textwidth]{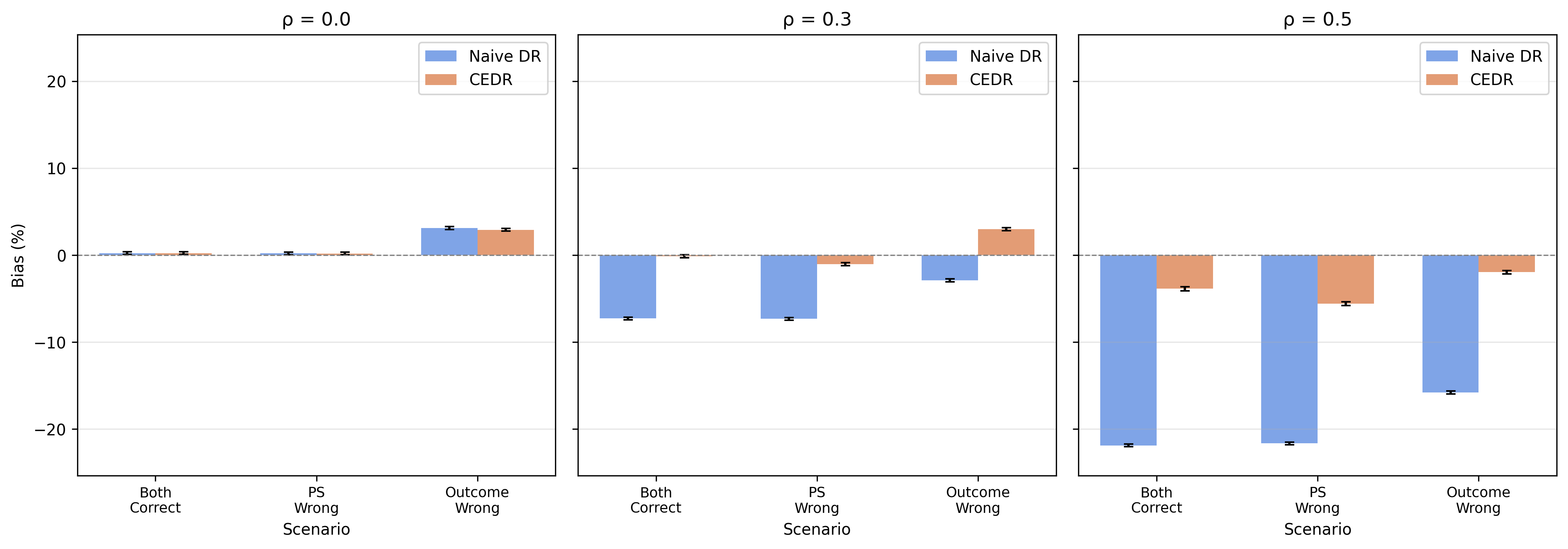}
    \caption{Bias (\%) by endogeneity level and misspecification  in scenario 1. }
    \label{fig:bias_single}
\end{figure*}

We present results for $n = 8000$ across all endogeneity levels and misspecification scenarios. Tables~\ref{tab:sim_single} and~\ref{tab:sim_dual} report bias and standard deviation with 95\% confidence intervals. Bias is expressed as a percentage of the true treatment effect (computed as $100 \times (\hat{\tau} - \tau) / \tau$). Figures~\ref{fig:bias_single} and~\ref{fig:bias_dual} provide a visual comparison. Additional experimental results can be found in the Appendix.

\subsubsection{No Endogeneity ($\rho = 0$)}

When there is no endogeneity, the naive DR and CEDR estimators perform nearly identically across all scenarios (Tables~\ref{tab:sim_single} -- \ref{tab:sim_dual}). Under correct specification of both models, bias is negligible for both estimators (below 0.25\%). The doubly robust property is observed when only the propensity score is misspecified (PS Wrong), bias remains near zero and when only the outcome model is misspecified (Outcome Wrong), moderate bias emerges (3 -- 4\%). Importantly, the CEDR does not introduce additional bias when endogeneity is absent.

\subsubsection{Moderate Endogeneity ($\rho = 0.3$)}

The benefits of the CEDR estimator become evident here. At $\rho = 0.3$ with both models correctly specified, the naive DR estimator exhibits $-7.3\%$ bias in Scenario 1 and $-9.7\%$ in Scenario 2. The CEDR estimator reduces this to $-0.1\%$ and $2.0\%$, respectively. Resulting in a bias reduction exceeding 80\% in both cases.

\subsubsection{Strong Endogeneity ($\rho = 0.5$ )}

At $\rho = 0.5$, the naive DR estimator is biased ($-21.9\%$ in Scenario 1, $-28.0\%$ in Scenario 2), while the CEDR estimator reduces bias to $-3.9\%$ and $-2.9\%$, respectively. Under propensity score misspecification (PS Wrong), the CEDR estimator maintains low bias ($-5.6\%$ Scenario 1, $-2.5\%$ Scenario 2) compared to the naive estimator ($-21.6\%$ Scenario 1, $-27.6\%$ Scenario 2). Under outcome model misspecification (Outcome Wrong), the CEDR estimator achieves $-2.0\%$ bias (Scenario 1) and $-1.2\%$ bias (Scenario 2), compared to $-15.8\%$ and $-23.2\%$ for the naive estimator.

\subsubsection{Variance--Bias Tradeoff}

The CEDR estimator exhibits slightly higher standard deviation than the naive DR estimator across all conditions. At $\rho = 0$, standard deviations are comparable (approximately 0.048 -- 0.053 for both). At $\rho = 0.5$, the CEDR standard deviation increases to 0.07 -- 0.11 compared to 0.04 -- 0.06 for the naive estimator. This modest increase in variance is outweighed by the  bias reduction at $\rho = 0.5$ with both models correct. 


\begin{table*}[t]
\centering
\caption{Monte Carlo results for Scenario 1.}
\label{tab:sim_single}
\footnotesize
\begin{tabular}{cl rr rr}
\toprule
& & \multicolumn{2}{c}{\textbf{Naive DR}} & \multicolumn{2}{c}{\textbf{CEDR}} \\
\cmidrule(lr){3-4} \cmidrule(lr){5-6}
$\rho$ & Scenario & Bias (\%) [95\% CI] & Std [95\% CI] & Bias (\%) [95\% CI] & Std [95\% CI] \\
\midrule
0 & Both Correct & 0.24 [0.08, 0.39] & 0.050 [0.048, 0.053] & 0.23 [0.07, 0.38] & 0.050 [0.048, 0.053] \\
  & PS Wrong & 0.21 [0.06, 0.35] & 0.048 [0.046, 0.050] & 0.19 [0.04, 0.34] & 0.049 [0.047, 0.051] \\
  & Outcome Wrong & 3.10 [2.94, 3.27] & 0.053 [0.051, 0.055] & 2.91 [2.75, 3.08] & 0.053 [0.050, 0.055] \\

\midrule
0.3 & Both Correct & $-$7.29 [$-$7.44, $-$7.13] & 0.050 [0.048, 0.052] & $-$0.13 [$-$0.31, 0.05] & 0.057 [0.055, 0.060] \\
    & PS Wrong & $-$7.32 [$-$7.48, $-$7.17] & 0.050 [0.048, 0.052] & $-$1.04 [$-$1.21, $-$0.87] & 0.054 [0.052, 0.057] \\
    & Outcome Wrong & $-$2.91 [$-$3.08, $-$2.74] & 0.055 [0.053, 0.058] & 2.99 [2.82, 3.17] & 0.057 [0.054, 0.059] \\

\midrule
0.5 & Both Correct & $-$21.87 [$-$22.02, $-$21.72] & 0.049 [0.047, 0.051] & $-$3.86 [$-$4.10, $-$3.61] & 0.079 [0.075, 0.082] \\
    & PS Wrong & $-$21.64 [$-$21.79, $-$21.49] & 0.049 [0.047, 0.051] & $-$5.60 [$-$5.82, $-$5.38] & 0.070 [0.067, 0.073] \\
    & Outcome Wrong & $-$15.78 [$-$15.95, $-$15.60] & 0.056 [0.054, 0.059] & $-$1.96 [$-$2.17, $-$1.76] & 0.066 [0.063, 0.069] \\

\bottomrule
\end{tabular}
\end{table*}

\begin{table*}[t]
\centering
\caption{Monte Carlo results for Scenario 2. }
\label{tab:sim_dual}
\footnotesize
\begin{tabular}{cl rr rr}
\toprule
& & \multicolumn{2}{c}{\textbf{Naive DR}} & \multicolumn{2}{c}{\textbf{CEDR}} \\
\cmidrule(lr){3-4} \cmidrule(lr){5-6}
$\rho$ & Scenario & Bias (\%) [95\% CI] & Std [95\% CI] & Bias (\%) [95\% CI] & Std [95\% CI] \\
\midrule
0 & Both Correct & $-$0.05 [$-$0.20, 0.10] & 0.048 [0.046, 0.050] & $-$0.05 [$-$0.21, 0.10] & 0.049 [0.047, 0.051] \\
  & PS Wrong & 0.00 [$-$0.15, 0.15] & 0.048 [0.046, 0.050] & 0.00 [$-$0.15, 0.15] & 0.049 [0.047, 0.051] \\
  & Outcome Wrong & 3.71 [3.55, 3.87] & 0.052 [0.049, 0.054] & 3.64 [3.47, 3.80] & 0.053 [0.051, 0.055] \\

\midrule
0.3 & Both Correct & $-$9.65 [$-$9.79, $-$9.52] & 0.044 [0.042, 0.046] & 1.96 [1.79, 2.14] & 0.057 [0.055, 0.060] \\
    & PS Wrong & $-$9.64 [$-$9.79, $-$9.50] & 0.046 [0.044, 0.048] & 1.59 [1.42, 1.76] & 0.056 [0.054, 0.059] \\
    & Outcome Wrong & $-$5.46 [$-$5.59, $-$5.32] & 0.044 [0.043, 0.047] & 5.19 [5.01, 5.36] & 0.057 [0.055, 0.060] \\

\midrule
0.5 & Both Correct & $-$27.98 [$-$28.12, $-$27.85] & 0.044 [0.042, 0.046] & $-$2.90 [$-$3.18, $-$2.62] & 0.090 [0.086, 0.094] \\
    & PS Wrong & $-$27.63 [$-$27.76, $-$27.50] & 0.042 [0.040, 0.044] & $-$2.54 [$-$2.80, $-$2.28] & 0.084 [0.080, 0.087] \\
    & Outcome Wrong & $-$23.18 [$-$23.33, $-$23.03] & 0.049 [0.046, 0.051] & $-$1.18 [$-$1.46, $-$0.91] & 0.090 [0.086, 0.094] \\

\bottomrule
\end{tabular}
\end{table*}


\begin{figure*}[t]
    \centering
    \includegraphics[width=\textwidth]{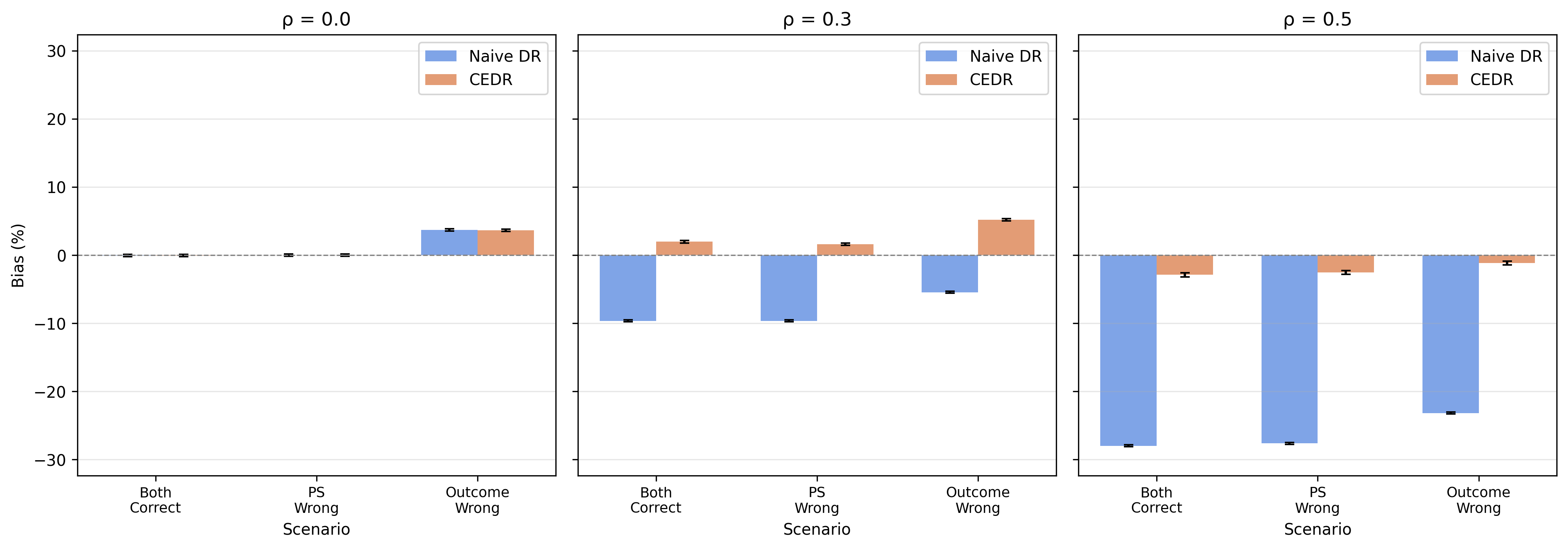}
    \caption{Bias (\%) by endogeneity level and misspecification in scenario 2. }
    \label{fig:bias_dual}
\end{figure*}

\section{Empirical Application: Nutrition Counseling and Blood Pressure} \label{sec4}

In this section, we apply the CEDR estimator to evaluate the causal effect of nutrition counseling from healthcare providers on  blood pressure (BP), using data from the National Health and Nutrition Examination Survey (NHANES).This application illustrates a setting where endogenous proxy covariates arise naturally. Income-to-poverty ratio serves as a proxy for socioeconomic resources and access to care. Similarly, BMI proxies for metabolic health. Both variables are endogenous, that CEDR is designed to address.

\subsection{Data Description}

We use publicly available data from the NHANES pre-pandemic cycle (2017--March 2020), a nationally representative cross-sectional survey administered by the National Center for Health Statistics. NHANES collects demographic, dietary, examination, and laboratory data through in-home interviews and standardized clinical assessments at mobile examination centers.

Our analysis sample comprises $n = 1{,}255$ adults aged 18 -- 85 with complete data on all variables. The binary treatment indicator is constructed from question DBQ700, which asks whether the participant received dietary advice from a healthcare provider in the preceding 12 months ($T_i = 1$ if yes, $T_i = 0$ if no). The primary outcome is blood pressure, measured as the average of up to three oscillometric readings taken during the clinical examination.

We designate two covariates as endogenous, the income-to-poverty ratio (INDFMPIR) and BMI (BMXBMI). Income reflects socioeconomic resources that correlate with unmeasured health behaviors and access to care, while BMI is influenced by unobserved genetic, metabolic, and lifestyle factors. Both variables are statistically non-normal (Anderson-Darling and Cram´er-von Mises tests $p < 0.001$ for each), satisfying the distributional requirement for copula identification. BMI exhibits pronounced right-skewness (skewness $= 1.33$), providing strong identification and income-to-poverty ratio, while non-normal, has weaker skewness ($-0.35$).

Sociodemographic variables include age, sex, education level, and race/ethnicity. We additionally condition on three confounders, smoking status, alcohol consumption, and self-reported diabetes diagnosis. A summary of participant characteristics is presented in Table~\ref{tab:nhanes_desc}.

\begin{table}[t]
\centering
\caption{Descriptive statistics for the NHANES analysis sample ($n = 1{,}255$). Continuous variables are reported as mean $\pm$ SD; binary variables as percentages.}\label{tab:nhanes_desc}
\begin{tabular*}{\textwidth}{@{\extracolsep\fill}lcccc}
\toprule
\textbf{Variable} & \textbf{Overall} & \textbf{Treated} ($n=294$) & \textbf{Control} ($n=961$)  \\
\midrule
Age (years) & $51.92 \pm 17.53$ & $54.58 \pm 17.32$ & $51.11 \pm 17.52$  \\
Income-poverty ratio & $3.29 \pm 1.65$ & $2.99 \pm 1.70$ & $3.38 \pm 1.62$  \\
BMI (kg/m$^2$) & $27.63 \pm 5.85$ & $26.85 \pm 5.30$ & $27.86 \pm 6.00$ \\
BP (mmHg) & $122.96 \pm 17.92$ & $126.77 \pm 19.51$ & $121.79 \pm 17.25$   \\
Alcohol (drinks/week) & $4.16 \pm 10.04$ & $4.91 \pm 15.47$ & $3.93 \pm 7.65$  \\
Male (\%) & $54.9$ & $60.5$ & $53.2$  \\
Current smoker (\%) & $13.9$ & $17.3$ & $12.9$ \\
Diabetes (\%) & $10.1$ & $11.6$ & $9.7$  \\
\bottomrule
\end{tabular*}
\end{table}

\subsection{Results}

We apply both the naive doubly robust (DR) estimator and the CEDR estimator, treating income-to-poverty ratio and BMI as endogenous covariates with Gaussian copula corrections. All remaining covariates enter both the propensity score and outcome models as exogenous regressors. Standard errors are computed via bootstrap with 5000 replications.

\begin{table}[t]
\centering
\caption{Estimated average treatment effect of nutrition counseling on blood pressure (mmHg).}\label{tab:nhanes_results}
\begin{tabular*}{\textwidth}{@{\extracolsep\fill}lccc}
\toprule
\textbf{Estimator} & \textbf{ATE (mmHg)}  & \textbf{ ATE 95\% CI} & \textbf{SE}\\
\midrule
Naive DR & $2.846$ & $[0.530, \; 5.342]$& $1.220$  \\
CEDR & $1.750$  & $[-6.988, \; 7.344]$& $3.743$ \\
\bottomrule
\end{tabular*}
\end{table}

From table~\ref{tab:nhanes_results} the naive DR estimator yields an ATE of $2.846$ mmHg (95\% CI: $[0.530, 5.342]$), suggesting that nutrition counseling is associated with \emph{higher} blood pressure. 
After applying the copula correction, the CEDR estimator reduces the point estimate to $1.750$ mmHg (95\% CI: $[-6.988, 7.344]$). The corrected estimate is not statistically significant, suggesting that much of the positive relation captured by the naive estimator results from the  bias coming from the endogenous covariates rather than a true causal effect of counseling on blood pressure.

The wider confidence interval for CEDR reflects the additional estimation uncertainty introduced by the copula correction terms, a variance--bias tradeoff that mirrors the patterns observed in our Monte Carlo experiments. This increased variance is expected when identification relies on distributional assumptions rather than exclusion restrictions.

\section{Discussion} \label{sec5}

To evaluate CEDR, we used synthetic data based on two designs outlined in Section~\ref{sec:simulations}. Each design represented varying strengths of endogeneity, with $\rho = 0$ and $\rho = 0.5$ representing no endogeneity and strong endogeneity. In the synthetic data, the true ATE ($\tau = 2$) and the sources of endogeneity were known, allowing us to directly assess bias reduction. Under two scenarios, our findings indicate that CEDR reduces bias by over 80\% relative to the naive DR estimator when at least one model is correctly specified, with a modest increase in variance.

We also validated the applicability of CEDR using NHANES data to estimate the causal effect of nutrition counseling on blood pressure. The naive DR estimator yielded a statistically significant increase in blood pressure of 2.846 mmHg. A finding that contradicts literature supporting the efficacy of dietary interventions for reducing blood pressure \citep{couch2021dietary, riegel2018efficacy}. After applying the copula correction, CEDR reduced this estimate to 1.750 mmHg, which was statistically insignificant. These results illustrate the value of CEDR in observational health research. The naive DR estimate would lead a policymaker to conclude that nutrition counseling is associated with a increase in blood pressure.  The CEDR estimate,  corrects the spurious positive association and produces results that does not go against the literature, where dietary counseling interventions typically yield modest reductions in blood pressure. 

The empirical application has several limitations. The NHANES data is cross-sectional, which makes establishing temporal ordering between counseling receipt and blood pressure measurement difficult. Treatment is self-reported, introducing the possibility of recall bias.  Additionally, copula identification relies on distributional non-normality of the endogenous covariates. While BMI provides strong identification through its pronounced right-skewness (1.33), the income-to-poverty ratio exhibits weaker skewness ($-0.35$). This contributes to the wider confidence intervals observed for CEDR relative to the naive estimator. Applications with multiple strongly skewed endogenous covariates would be expected to yield tighter intervals.

More broadly, CEDR requires that the Gaussian copula correctly specifies the dependence structure between the endogenous covariates and the structural error. While this is a standard assumption in the copula-based endogeneity literature \citep{park2012handling, liengaard2025dealing}, misspecification of the copula family would compromise the correction. The method also cannot identify endogeneity when all endogenous covariates are normally distributed, as the copula term becomes collinear with the covariate itself. These constraints define the boundary conditions under which CEDR is applicable and should guide practitioners in assessing whether the method is appropriate for their setting.

Given these limitations, we caution against drawing definitive clinical conclusions from the empirical application alone. The primary purpose of the NHANES analysis is to demonstrate the practical use and value of CEDR in a real-world setting. To draw robust policy conclusions, researchers should apply CEDR to multiple high-quality datasets and, where possible, benchmark results against experimental evidence.

For future research, we plan to extend CEDR in several directions. First, generalizing the framework to continuous and multi-valued treatments, where covariate endogeneity poses similar challenges but the doubly robust estimator requires reformulation beyond the binary treatment setting. Second, extending the copula correction to accommodate categorical endogenous covariates.

\section{Conclusion} \label{sec6}

In this paper, we introduce the Copula-corrected Endogeneity-adjusted Doubly Robust (CEDR) estimator for causal inference when proxy covariates are endogenous and instrumental variables are unavailable. CEDR addresses this by integrating Gaussian copula corrections into the DR estimator, and exploits the non-normality of endogenous covariates for identification. We establish that CEDR preserves the double robustness property. The estimator is consistent for the average treatment effect when either the copula-corrected propensity score or the copula-corrected outcome model is correctly specified.

Monte Carlo simulations confirm that CEDR reduces bias across a range of endogeneity strengths and misspecification scenarios, with only small increases in variance. When endogeneity is absent, CEDR is innocuous, performing comparably to the naive DR estimator. In the NHANES application, CEDR attenuates a spurious positive association between nutrition counseling and blood pressure, producing results more consistent with the literature.

However, CEDR is currently limited to settings with binary treatment, cross-sectional data, and endogenous covariates that exhibit sufficient non-normality for copula identification. The method requires correct specification of the Gaussian copula dependence structure and cannot address endogeneity arising from normally distributed covariates. Despite these constraints, CEDR fills a gap in the causal inference toolkit by providing a practical, IV-free approach to endogeneity that maintains robustness to model misspecification. Our analysis suggests that standard observational estimates of treatment effects in healthcare settings may be substantially biased when proxy covariates are endogenous, and that copula-based corrections offer a viable path toward more reliable causal conclusions.

\section*{Declaration}

The authors report there are no competing interests to declare.

\section*{Consent and Approval Statement}

Not Applicable, since our study utilizes publicly available data. 

\section*{Funding}

Financial support from the National Science Foundation (Award Number:
2047094) is greatly acknowledged.

\bibliographystyle{apalike} 
\bibliography{bib}

\appendix
\section{Additional Numerical Experiments}\label{appendix}

This appendix reports the complete Monte Carlo simulation results for $n  = (2{,}000,\; 4{,}000,\; 16{,}000,\; 32{,}000)$ across all endogeneity levels $\rho =  (0.0, 0.3, 0.50)$ and all three misspecification scenarios (Both Correct, PS Wrong, Outcome Wrong). Bias is expressed as a percentage of the true treatment effect ($\tau = 2$), and 95\% confidence intervals are reported in brackets.

The results across all sample sizes are consistent with the findings reported in the main manuscript. When endogeneity is absent ($\rho = 0$), the naive DR and CEDR estimators perform comparably, confirming that the copula correction does not introduce spurious bias. Under moderate ($\rho = 0.3$) and strong ($\rho = 0.5$) endogeneity, CEDR achieves substantial bias reduction relative to the naive estimator across all sample sizes. The doubly robust property is preserved under model misspecification. As expected, when we increase the sample size both bias and variance decrease, and the CEDR exhibits modestly higher standard deviations than the naive estimator, reflecting the variance -- bias tradeoff discussed in Section~3.2.4. 


\subsection*{A.1. Scenario 1: Complete Results}

\begin{table}[htbp]
\centering
\caption{Scenario 1, $n = 2{,}000$: Bias (\%) and standard deviation.}
\label{tab:s1_n2000}
\resizebox{\columnwidth}{!}{%
\begin{tabular}{cl rr rr}
\hline
& & \multicolumn{2}{c}{Naive DR} & \multicolumn{2}{c}{CEDR} \\
\cmidrule(lr){3-4} \cmidrule(lr){5-6}
$\rho$ & Scenario & Bias (\%) [95\% CI] & Std [95\% CI] & Bias (\%) [95\% CI] & Std [95\% CI] \\
\hline
0 & Both Correct & 0.06 [$-$0.26, 0.37] & 0.103 [0.099, 0.108] & 0.08 [$-$0.25, 0.42] & 0.108 [0.103, 0.112] \\
  & PS Wrong & 0.12 [$-$0.19, 0.42] & 0.097 [0.093, 0.102] & 0.14 [$-$0.18, 0.46] & 0.102 [0.098, 0.107] \\
  & Outcome Wrong & 2.93 [2.59, 3.27] & 0.110 [0.106, 0.116] & 2.61 [2.25, 2.97] & 0.117 [0.112, 0.122] \\
\hline
0.3 & Both Correct & $-$7.23 [$-$7.55, $-$6.91] & 0.104 [0.100, 0.109] & 0.14 [$-$0.23, 0.51] & 0.119 [0.114, 0.125] \\
    & PS Wrong & $-$7.33 [$-$7.65, $-$7.02] & 0.101 [0.097, 0.106] & $-$0.93 [$-$1.30, $-$0.57] & 0.116 [0.111, 0.122] \\
    & Outcome Wrong & $-$2.89 [$-$3.24, $-$2.54] & 0.112 [0.107, 0.117] & 2.89 [2.52, 3.25] & 0.118 [0.114, 0.124] \\
\hline
0.5 & Both Correct & $-$22.08 [$-$22.38, $-$21.78] & 0.097 [0.093, 0.101] & $-$3.05 [$-$3.59, $-$2.50] & 0.177 [0.170, 0.185] \\
    & PS Wrong & $-$21.79 [$-$22.09, $-$21.49] & 0.097 [0.093, 0.101] & $-$5.32 [$-$5.77, $-$4.86] & 0.148 [0.142, 0.155] \\
    & Outcome Wrong & $-$15.88 [$-$16.22, $-$15.54] & 0.110 [0.105, 0.115] & $-$2.10 [$-$2.52, $-$1.68] & 0.135 [0.129, 0.141] \\
\hline
\end{tabular}%
}
\end{table}

\begin{table}[htbp]
\centering
\caption{Scenario 1, $n = 4{,}000$: Bias (\%) and standard deviation.}
\label{tab:s1_n4000}
\resizebox{\columnwidth}{!}{%
\begin{tabular}{cl rr rr}
\hline
& & \multicolumn{2}{c}{Naive DR} & \multicolumn{2}{c}{CEDR} \\
\cmidrule(lr){3-4} \cmidrule(lr){5-6}
$\rho$ & Scenario & Bias (\%) [95\% CI] & Std [95\% CI] & Bias (\%) [95\% CI] & Std [95\% CI] \\
\hline
0 & Both Correct & $-$0.06 [$-$0.28, 0.17] & 0.073 [0.070, 0.076] & $-$0.11 [$-$0.34, 0.12] & 0.075 [0.071, 0.078] \\
  & PS Wrong & $-$0.11 [$-$0.33, 0.10] & 0.069 [0.066, 0.072] & $-$0.16 [$-$0.38, 0.06] & 0.071 [0.068, 0.074] \\
  & Outcome Wrong & 2.83 [2.60, 3.07] & 0.077 [0.074, 0.081] & 2.51 [2.27, 2.76] & 0.080 [0.076, 0.083] \\
\hline
0.3 & Both Correct & $-$7.20 [$-$7.43, $-$6.97] & 0.075 [0.072, 0.078] & 0.01 [$-$0.24, 0.26] & 0.082 [0.078, 0.086] \\
    & PS Wrong & $-$7.30 [$-$7.53, $-$7.07] & 0.074 [0.071, 0.078] & $-$1.05 [$-$1.30, $-$0.81] & 0.079 [0.076, 0.083] \\
    & Outcome Wrong & $-$2.77 [$-$3.03, $-$2.52] & 0.082 [0.079, 0.086] & 3.09 [2.83, 3.34] & 0.081 [0.078, 0.085] \\
\hline
0.5 & Both Correct & $-$21.89 [$-$22.10, $-$21.68] & 0.069 [0.066, 0.072] & $-$3.31 [$-$3.66, $-$2.96] & 0.112 [0.107, 0.117] \\
    & PS Wrong & $-$21.62 [$-$21.83, $-$21.41] & 0.069 [0.066, 0.072] & $-$5.23 [$-$5.54, $-$4.91] & 0.101 [0.097, 0.105] \\
    & Outcome Wrong & $-$15.75 [$-$15.99, $-$15.52] & 0.077 [0.074, 0.080] & $-$1.75 [$-$2.03, $-$1.47] & 0.089 [0.086, 0.094] \\
\hline
\end{tabular}%
}
\end{table}

\begin{table}[htbp]
\centering
\caption{Scenario 1, $n = 16{,}000$: Bias (\%) and standard deviation.}
\label{tab:s1_n16000}
\resizebox{\columnwidth}{!}{%
\begin{tabular}{cl rr rr}
\hline
& & \multicolumn{2}{c}{Naive DR} & \multicolumn{2}{c}{CEDR} \\
\cmidrule(lr){3-4} \cmidrule(lr){5-6}
$\rho$ & Scenario & Bias (\%) [95\% CI] & Std [95\% CI] & Bias (\%) [95\% CI] & Std [95\% CI] \\
\hline
0 & Both Correct & $-$0.14 [$-$0.25, $-$0.03] & 0.036 [0.035, 0.038] & $-$0.15 [$-$0.26, $-$0.04] & 0.037 [0.035, 0.038] \\
  & PS Wrong & $-$0.09 [$-$0.20, 0.01] & 0.035 [0.033, 0.036] & $-$0.10 [$-$0.21, 0.01] & 0.035 [0.034, 0.037] \\
  & Outcome Wrong & 2.68 [2.57, 2.80] & 0.038 [0.037, 0.040] & 2.53 [2.41, 2.65] & 0.039 [0.037, 0.040] \\
\hline
0.3 & Both Correct & $-$7.31 [$-$7.43, $-$7.20] & 0.036 [0.035, 0.038] & $-$0.24 [$-$0.36, $-$0.12] & 0.039 [0.038, 0.041] \\
    & PS Wrong & $-$7.30 [$-$7.40, $-$7.19] & 0.034 [0.033, 0.036] & $-$1.09 [$-$1.21, $-$0.98] & 0.037 [0.035, 0.039] \\
    & Outcome Wrong & $-$2.99 [$-$3.11, $-$2.87] & 0.039 [0.037, 0.041] & 2.93 [2.81, 3.05] & 0.039 [0.037, 0.040] \\
\hline
0.5 & Both Correct & $-$21.88 [$-$21.99, $-$21.78] & 0.034 [0.032, 0.035] & $-$3.83 [$-$3.99, $-$3.67] & 0.052 [0.050, 0.055] \\
    & PS Wrong & $-$21.60 [$-$21.71, $-$21.50] & 0.034 [0.033, 0.036] & $-$5.51 [$-$5.66, $-$5.36] & 0.049 [0.047, 0.051] \\
    & Outcome Wrong & $-$15.75 [$-$15.87, $-$15.64] & 0.038 [0.036, 0.039] & $-$1.78 [$-$1.92, $-$1.65] & 0.043 [0.041, 0.045] \\
\hline
\end{tabular}%
}
\end{table}

\begin{table}[htbp]
\centering
\caption{Scenario 1, $n = 32{,}000$: Bias (\%) and standard deviation.}
\label{tab:s1_n32000}
\resizebox{\columnwidth}{!}{%
\begin{tabular}{cl rr rr}
\hline
& & \multicolumn{2}{c}{Naive DR} & \multicolumn{2}{c}{CEDR} \\
\cmidrule(lr){3-4} \cmidrule(lr){5-6}
$\rho$ & Scenario & Bias (\%) [95\% CI] & Std [95\% CI] & Bias (\%) [95\% CI] & Std [95\% CI] \\
\hline
0 & Both Correct & 0.02 [$-$0.05, 0.10] & 0.025 [0.024, 0.026] & 0.03 [$-$0.05, 0.11] & 0.025 [0.024, 0.026] \\
  & PS Wrong & 0.01 [$-$0.07, 0.08] & 0.024 [0.023, 0.025] & 0.01 [$-$0.07, 0.08] & 0.024 [0.023, 0.025] \\
  & Outcome Wrong & 2.84 [2.76, 2.92] & 0.026 [0.025, 0.027] & 2.71 [2.63, 2.79] & 0.026 [0.025, 0.027] \\
\hline
0.3 & Both Correct & $-$7.22 [$-$7.30, $-$7.14] & 0.026 [0.025, 0.027] & $-$0.11 [$-$0.20, $-$0.02] & 0.028 [0.027, 0.029] \\
    & PS Wrong & $-$7.22 [$-$7.29, $-$7.14] & 0.026 [0.025, 0.027] & $-$0.98 [$-$1.06, $-$0.90] & 0.027 [0.026, 0.028] \\
    & Outcome Wrong & $-$2.88 [$-$2.96, $-$2.79] & 0.028 [0.027, 0.029] & 3.07 [2.99, 3.16] & 0.028 [0.027, 0.029] \\
\hline
0.5 & Both Correct & $-$21.92 [$-$21.99, $-$21.84] & 0.025 [0.024, 0.026] & $-$3.87 [$-$3.98, $-$3.75] & 0.037 [0.035, 0.039] \\
    & PS Wrong & $-$21.63 [$-$21.70, $-$21.56] & 0.024 [0.023, 0.025] & $-$5.54 [$-$5.64, $-$5.44] & 0.034 [0.032, 0.035] \\
    & Outcome Wrong & $-$15.79 [$-$15.88, $-$15.70] & 0.028 [0.027, 0.030] & $-$1.79 [$-$1.88, $-$1.70] & 0.030 [0.029, 0.032] \\
\hline
\end{tabular}%
}
\end{table}

\subsubsection*{A.1.1. Figures for Scenario 1}


\begin{figure}[htbp]
\centering
\includegraphics[width=\columnwidth]{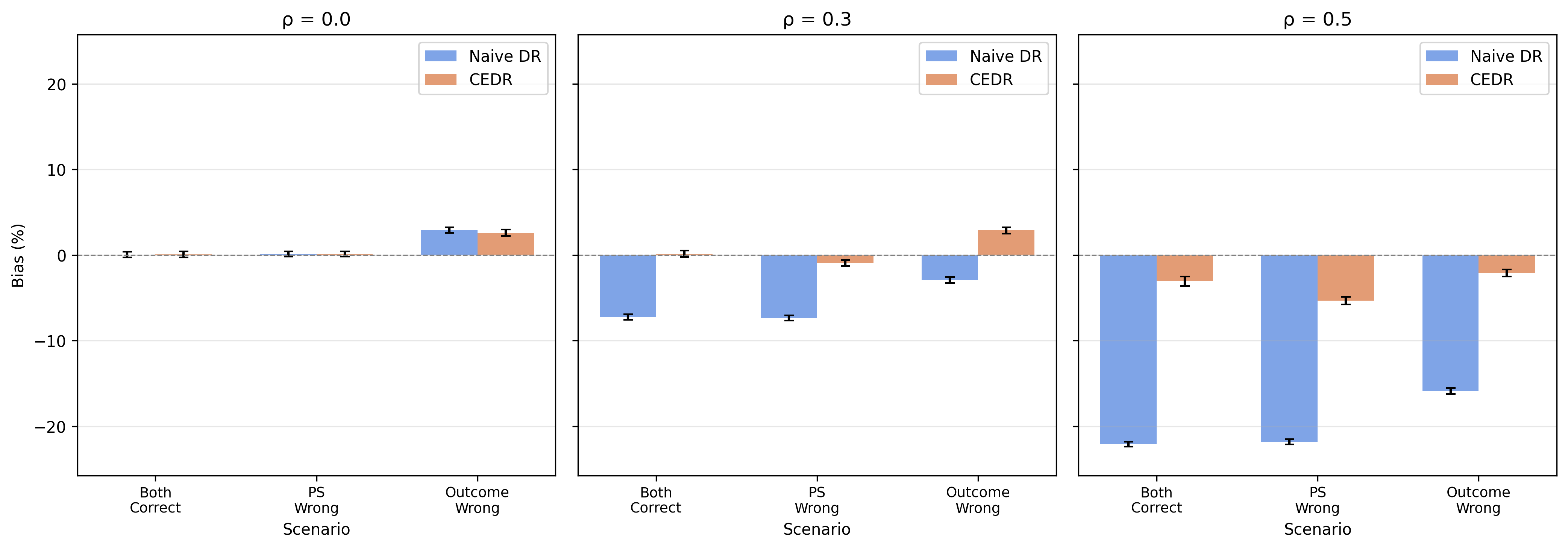}
\caption{Scenario 1, $n = 2{,}000$: Bias (\%) by endogeneity level and misspecification.}
\label{fig:s1_n2000}
\end{figure}

\begin{figure}[htbp]
\centering
\includegraphics[width=\columnwidth]{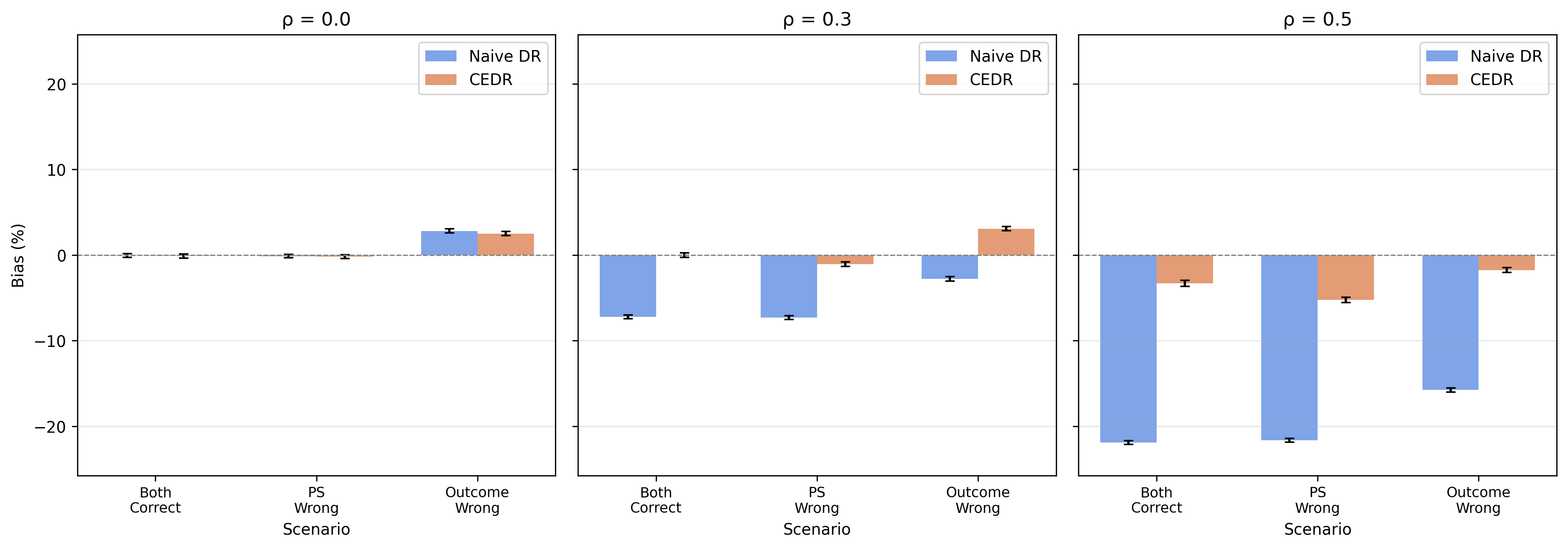}
\caption{Scenario 1, $n = 4{,}000$: Bias (\%) by endogeneity level and misspecification.}
\label{fig:s1_n4000}
\end{figure}

\begin{figure}[htbp]
\centering
\includegraphics[width=\columnwidth]{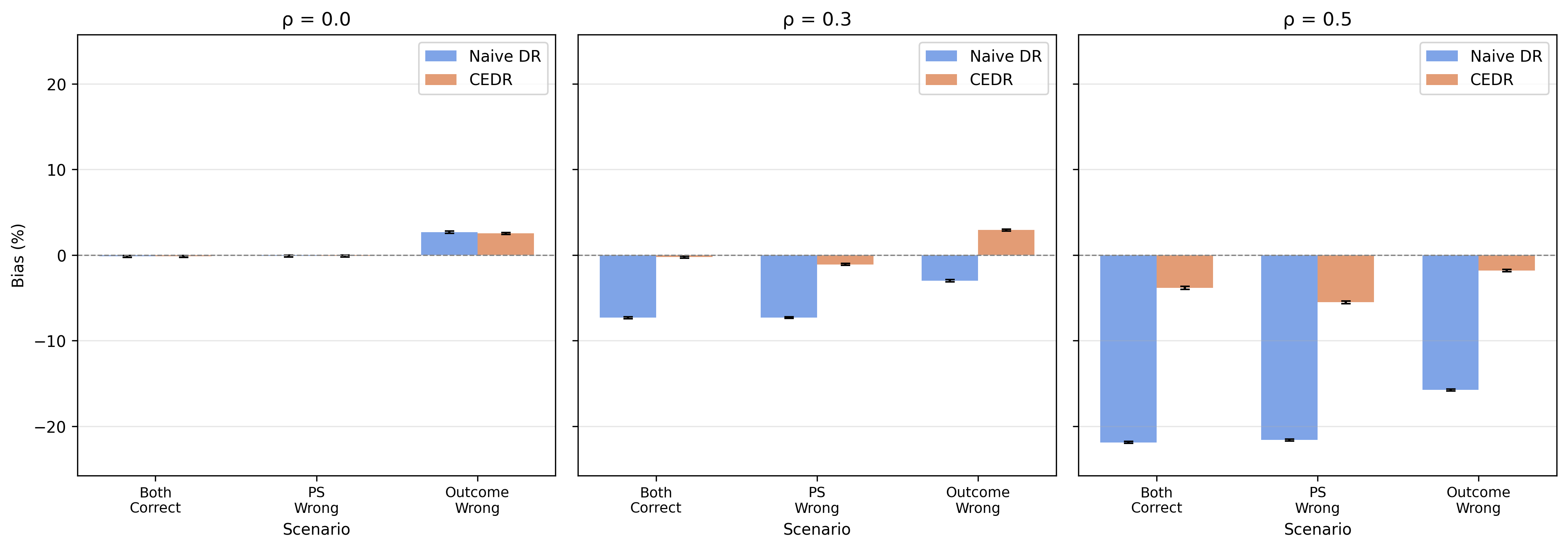}
\caption{Scenario 1, $n = 16{,}000$: Bias (\%) by endogeneity level and misspecification.}
\label{fig:s1_n16000}
\end{figure}

\begin{figure}[htbp]
\centering
\includegraphics[width=\columnwidth]{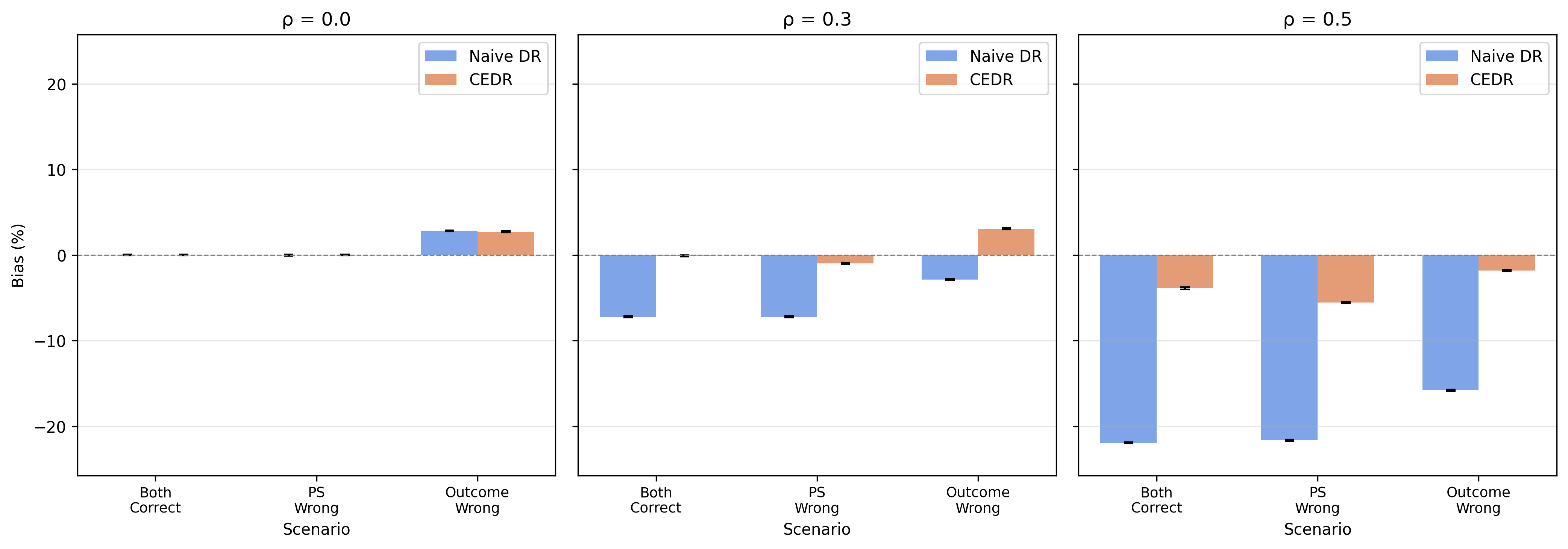}
\caption{Scenario 1, $n = 32{,}000$: Bias (\%) by endogeneity level and misspecification.}
\label{fig:s1_n32000}
\end{figure}


\subsection*{A.2. Scenario 2: Complete Results}

\begin{table}[htbp]
\centering
\caption{Scenario 2, $n = 2{,}000$: Bias (\%) and standard deviation.}
\label{tab:s2_n2000}
\resizebox{\columnwidth}{!}{%
\begin{tabular}{cl rr rr}
\hline
& & \multicolumn{2}{c}{Naive DR} & \multicolumn{2}{c}{CEDR} \\
\cmidrule(lr){3-4} \cmidrule(lr){5-6}
$\rho$ & Scenario & Bias (\%) [95\% CI] & Std [95\% CI] & Bias (\%) [95\% CI] & Std [95\% CI] \\
\hline
0 & Both Correct & 0.13 [$-$0.15, 0.40] & 0.089 [0.085, 0.093] & 0.14 [$-$0.15, 0.44] & 0.095 [0.091, 0.100] \\
  & PS Wrong & 0.43 [0.12, 0.75] & 0.102 [0.097, 0.106] & 0.41 [0.08, 0.75] & 0.108 [0.103, 0.113] \\
  & Outcome Wrong & 3.69 [3.39, 3.99] & 0.096 [0.092, 0.101] & 3.58 [3.26, 3.89] & 0.102 [0.098, 0.107] \\
\hline
0.3 & Both Correct & $-$9.92 [$-$10.18, $-$9.67] & 0.083 [0.080, 0.087] & 2.20 [1.77, 2.63] & 0.140 [0.134, 0.146] \\
    & PS Wrong & $-$10.10 [$-$10.36, $-$9.83] & 0.085 [0.081, 0.088] & 1.73 [1.33, 2.14] & 0.131 [0.126, 0.137] \\
    & Outcome Wrong & $-$5.70 [$-$5.97, $-$5.42] & 0.089 [0.085, 0.093] & 5.32 [4.91, 5.74] & 0.134 [0.128, 0.140] \\
\hline
0.5 & Both Correct & $-$27.83 [$-$28.11, $-$27.55] & 0.090 [0.087, 0.095] & $-$1.02 [$-$1.67, $-$0.37] & 0.209 [0.200, 0.218] \\
    & PS Wrong & $-$27.56 [$-$27.83, $-$27.29] & 0.086 [0.082, 0.090] & $-$1.13 [$-$1.71, $-$0.54] & 0.190 [0.182, 0.199] \\
    & Outcome Wrong & $-$23.03 [$-$23.34, $-$22.72] & 0.099 [0.095, 0.103] & 0.17 [$-$0.47, 0.81] & 0.207 [0.198, 0.216] \\
\hline
\end{tabular}%
}
\end{table}

\begin{table}[htbp]
\centering
\caption{Scenario 2, $n = 4{,}000$: Bias (\%) and standard deviation.}
\label{tab:s2_n4000}
\resizebox{\columnwidth}{!}{%
\begin{tabular}{cl rr rr}
\hline
& & \multicolumn{2}{c}{Naive DR} & \multicolumn{2}{c}{CEDR} \\
\cmidrule(lr){3-4} \cmidrule(lr){5-6}
$\rho$ & Scenario & Bias (\%) [95\% CI] & Std [95\% CI] & Bias (\%) [95\% CI] & Std [95\% CI] \\
\hline
0 & Both Correct & 0.07 [$-$0.14, 0.28] & 0.068 [0.065, 0.071] & 0.07 [$-$0.15, 0.29] & 0.071 [0.068, 0.074] \\
  & PS Wrong & 0.05 [$-$0.16, 0.26] & 0.068 [0.066, 0.071] & 0.04 [$-$0.18, 0.26] & 0.071 [0.068, 0.075] \\
  & Outcome Wrong & 3.84 [3.61, 4.06] & 0.072 [0.069, 0.075] & 3.75 [3.52, 3.99] & 0.075 [0.072, 0.079] \\
\hline
0.3 & Both Correct & $-$9.80 [$-$9.99, $-$9.60] & 0.062 [0.060, 0.065] & 1.65 [1.38, 1.92] & 0.087 [0.083, 0.091] \\
    & PS Wrong & $-$9.68 [$-$9.87, $-$9.48] & 0.062 [0.060, 0.065] & 1.52 [1.25, 1.79] & 0.088 [0.084, 0.092] \\
    & Outcome Wrong & $-$5.61 [$-$5.81, $-$5.40] & 0.067 [0.064, 0.070] & 4.94 [4.67, 5.21] & 0.088 [0.084, 0.092] \\
\hline
0.5 & Both Correct & $-$28.09 [$-$28.29, $-$27.89] & 0.064 [0.061, 0.067] & $-$2.52 [$-$2.96, $-$2.08] & 0.142 [0.136, 0.148] \\
    & PS Wrong & $-$27.87 [$-$28.06, $-$27.68] & 0.061 [0.058, 0.063] & $-$2.48 [$-$2.87, $-$2.10] & 0.124 [0.119, 0.130] \\
    & Outcome Wrong & $-$23.29 [$-$23.50, $-$23.07] & 0.069 [0.067, 0.073] & $-$1.01 [$-$1.43, $-$0.58] & 0.138 [0.132, 0.144] \\
\hline
\end{tabular}%
}
\end{table}

\begin{table}[htbp]
\centering
\caption{Scenario 2, $n = 16{,}000$: Bias (\%) and standard deviation.}
\label{tab:s2_n16000}
\resizebox{\columnwidth}{!}{%
\begin{tabular}{cl rr rr}
\hline
& & \multicolumn{2}{c}{Naive DR} & \multicolumn{2}{c}{CEDR} \\
\cmidrule(lr){3-4} \cmidrule(lr){5-6}
$\rho$ & Scenario & Bias (\%) [95\% CI] & Std [95\% CI] & Bias (\%) [95\% CI] & Std [95\% CI] \\
\hline
0 & Both Correct & 0.00 [$-$0.10, 0.11] & 0.033 [0.032, 0.035] & $-$0.00 [$-$0.11, 0.10] & 0.034 [0.033, 0.036] \\
  & PS Wrong & $-$0.01 [$-$0.12, 0.09] & 0.034 [0.032, 0.035] & $-$0.02 [$-$0.13, 0.08] & 0.035 [0.033, 0.036] \\
  & Outcome Wrong & 3.73 [3.62, 3.84] & 0.036 [0.034, 0.037] & 3.69 [3.58, 3.80] & 0.036 [0.035, 0.038] \\
\hline
0.3 & Both Correct & $-$9.78 [$-$9.88, $-$9.69] & 0.032 [0.030, 0.033] & 1.73 [1.60, 1.86] & 0.043 [0.041, 0.045] \\
    & PS Wrong & $-$9.65 [$-$9.75, $-$9.56] & 0.031 [0.030, 0.033] & 1.55 [1.42, 1.68] & 0.041 [0.040, 0.043] \\
    & Outcome Wrong & $-$5.61 [$-$5.71, $-$5.50] & 0.034 [0.033, 0.036] & 5.06 [4.92, 5.19] & 0.043 [0.042, 0.045] \\
\hline
0.5 & Both Correct & $-$28.20 [$-$28.29, $-$28.10] & 0.031 [0.030, 0.033] & $-$2.91 [$-$3.11, $-$2.71] & 0.064 [0.061, 0.067] \\
    & PS Wrong & $-$27.94 [$-$28.04, $-$27.84] & 0.031 [0.030, 0.033] & $-$2.70 [$-$2.88, $-$2.52] & 0.058 [0.056, 0.061] \\
    & Outcome Wrong & $-$23.38 [$-$23.48, $-$23.27] & 0.034 [0.033, 0.036] & $-$1.25 [$-$1.44, $-$1.06] & 0.062 [0.059, 0.064] \\
\hline
\end{tabular}%
}
\end{table}

\begin{table}[htbp]
\centering
\caption{Scenario 2, $n = 32{,}000$: Bias (\%) and standard deviation.}
\label{tab:s2_n32000}
\resizebox{\columnwidth}{!}{%
\begin{tabular}{cl rr rr}
\hline
& & \multicolumn{2}{c}{Naive DR} & \multicolumn{2}{c}{CEDR} \\
\cmidrule(lr){3-4} \cmidrule(lr){5-6}
$\rho$ & Scenario & Bias (\%) [95\% CI] & Std [95\% CI] & Bias (\%) [95\% CI] & Std [95\% CI] \\
\hline
0 & Both Correct & $-$0.06 [$-$0.13, 0.02] & 0.024 [0.023, 0.025] & $-$0.07 [$-$0.15, 0.00] & 0.025 [0.024, 0.026] \\
  & PS Wrong & $-$0.04 [$-$0.11, 0.04] & 0.025 [0.023, 0.026] & $-$0.04 [$-$0.12, 0.03] & 0.025 [0.024, 0.026] \\
  & Outcome Wrong & 3.66 [3.59, 3.74] & 0.026 [0.025, 0.027] & 3.62 [3.54, 3.71] & 0.026 [0.025, 0.028] \\
\hline
0.3 & Both Correct & $-$9.82 [$-$9.88, $-$9.75] & 0.022 [0.021, 0.023] & 1.59 [1.51, 1.68] & 0.027 [0.025, 0.028] \\
    & PS Wrong & $-$9.74 [$-$9.80, $-$9.67] & 0.021 [0.020, 0.022] & 1.45 [1.37, 1.54] & 0.027 [0.026, 0.028] \\
    & Outcome Wrong & $-$5.60 [$-$5.67, $-$5.53] & 0.024 [0.023, 0.025] & 5.00 [4.91, 5.09] & 0.029 [0.027, 0.030] \\
\hline
0.5 & Both Correct & $-$28.23 [$-$28.30, $-$28.16] & 0.023 [0.022, 0.024] & $-$3.05 [$-$3.19, $-$2.90] & 0.048 [0.046, 0.050] \\
    & PS Wrong & $-$27.99 [$-$28.06, $-$27.92] & 0.022 [0.021, 0.023] & $-$2.85 [$-$2.98, $-$2.72] & 0.042 [0.040, 0.044] \\
    & Outcome Wrong & $-$23.39 [$-$23.47, $-$23.32] & 0.025 [0.024, 0.026] & $-$1.31 [$-$1.45, $-$1.16] & 0.046 [0.044, 0.048] \\
\hline
\end{tabular}%
}
\end{table}

\subsubsection*{A.2.1. Figures for Scenario 2}


\begin{figure}[htbp]
\centering
\includegraphics[width=\columnwidth]{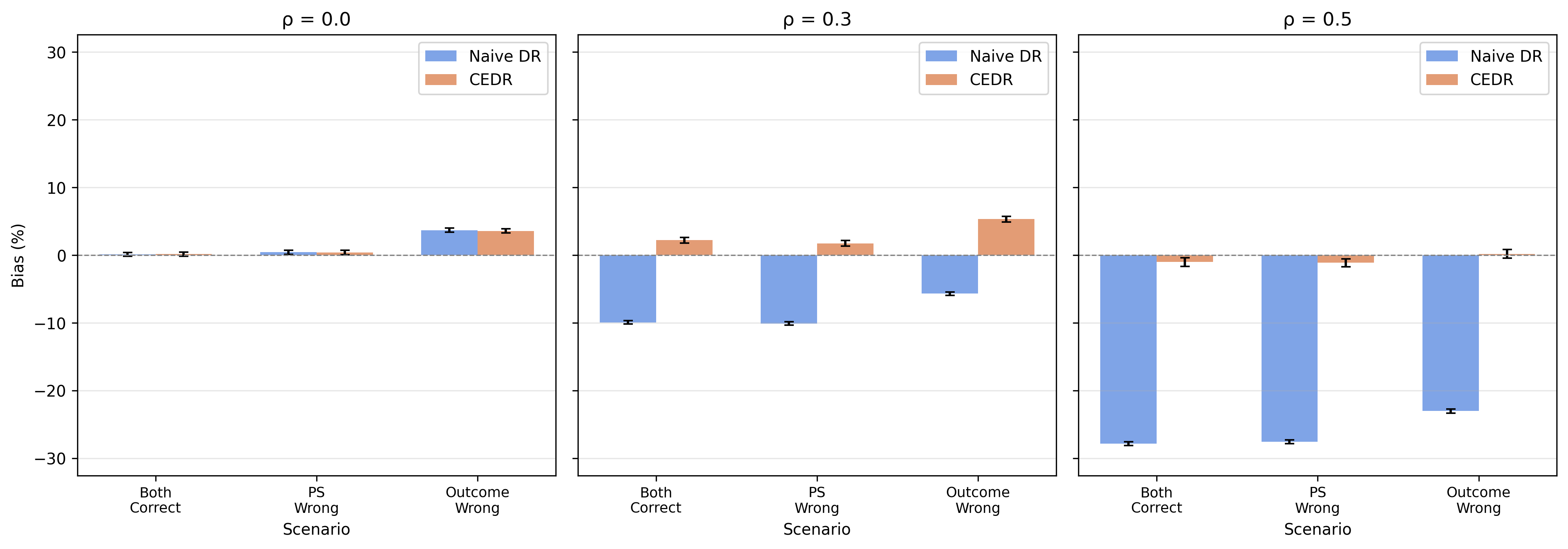}
\caption{Scenario 2, $n = 2{,}000$: Bias (\%) by endogeneity level and misspecification.}
\label{fig:s2_n2000}
\end{figure}

\begin{figure}[htbp]
\centering
\includegraphics[width=\columnwidth]{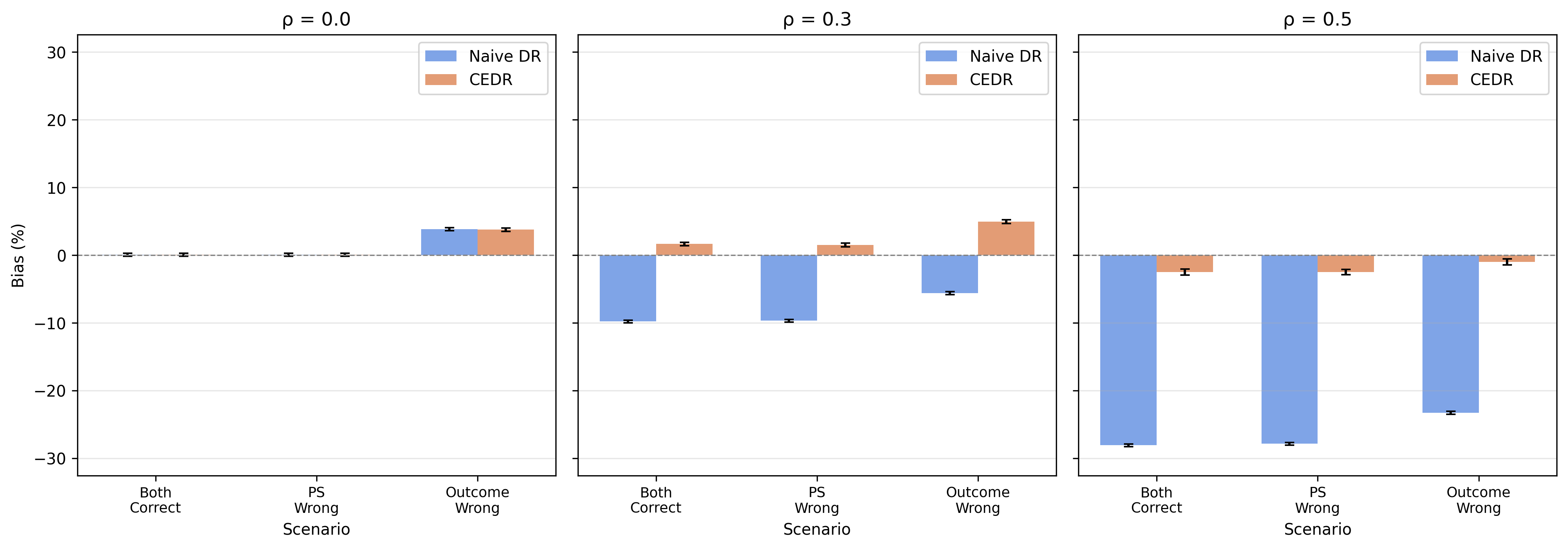}
\caption{Scenario 2, $n = 4{,}000$: Bias (\%) by endogeneity level and misspecification.}
\label{fig:s2_n4000}
\end{figure}

\begin{figure}[htbp]
\centering
\includegraphics[width=\columnwidth]{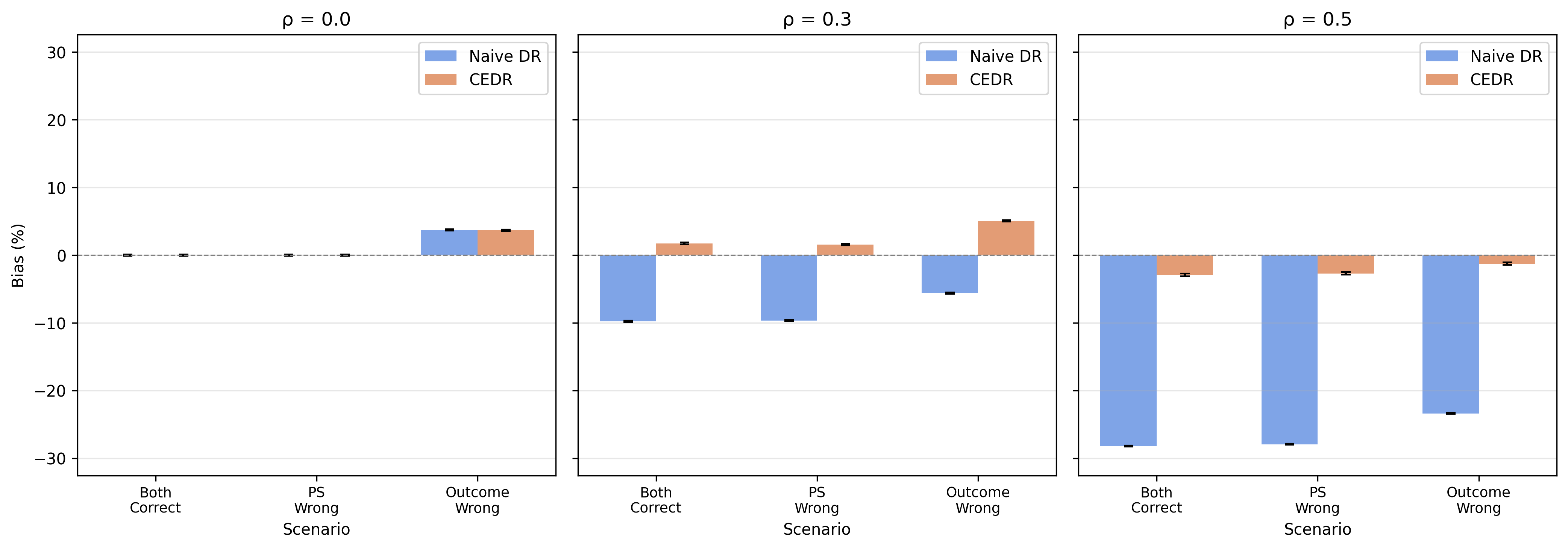}
\caption{Scenario 2, $n = 16{,}000$: Bias (\%) by endogeneity level and misspecification.}
\label{fig:s2_n16000}
\end{figure}

\begin{figure}[htbp]
\centering
\includegraphics[width=\columnwidth]{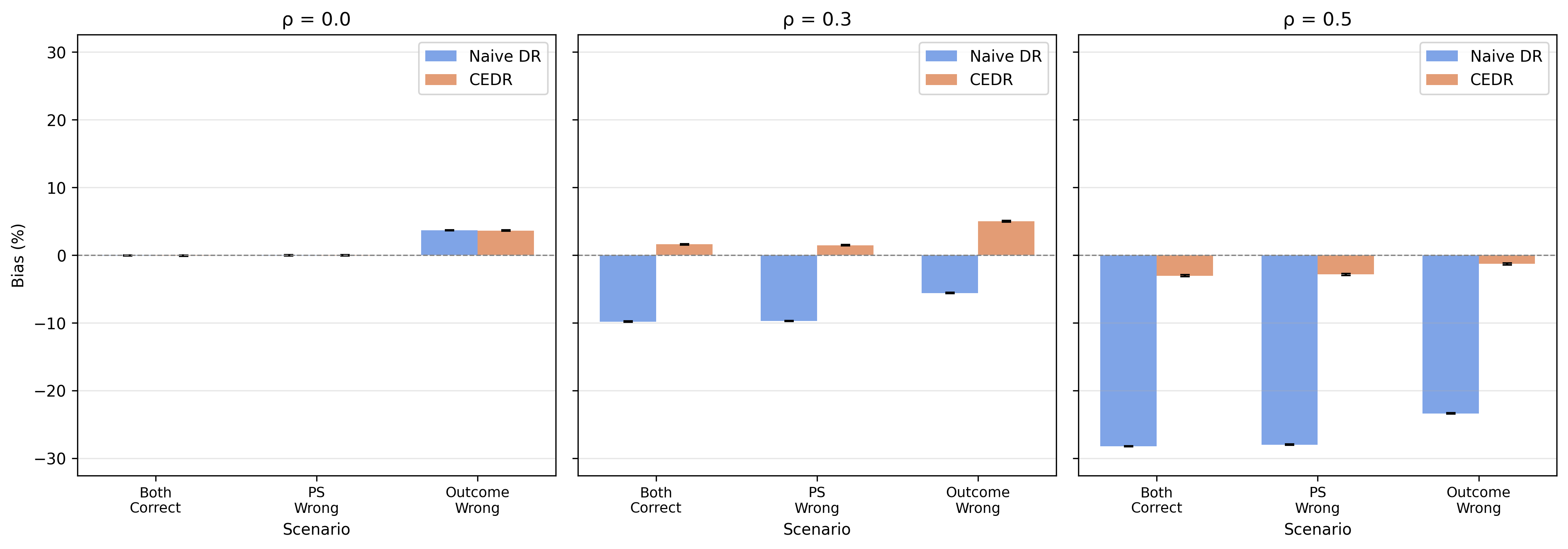}
\caption{Scenario 2, $n = 32{,}000$: Bias (\%) by endogeneity level and misspecification.}
\label{fig:s2_n32000}
\end{figure}

\end{document}